\documentclass[twocolumn,showpacs,superscriptaddress,amsmath,amssymb]{revtex4}
\usepackage{graphicx}
\usepackage{dcolumn}
\usepackage{bm}

\def\bd{
\begin{document}} \def\ed{\end{document}}
\def\bmp{\begin{minipage}} \def\emp{\end{minipage}}
\def\bcc{\begin{center}} \def\ecc{\end{center}}     \def\npg{\newpage}
\def\beq{\begin{equation}} \def\eeq{\end{equation}} \def\hph{\hphantom}
\def\be{\begin{equation}} \def\ee{\end{equation}} \def\r#1{$^{[#1]}$}
\def\n{\noindent} \def\ni{\noindent} \def\pa{\parindent}
\def\hs{\hskip} \def\vs{\vskip} \def\hf{\hfill} \def\ej{\vfill\eject}
\def\cl{\centerline} \def\ob{\obeylines}  \def\ls{\leftskip}
\def\underbar#1{$\setbox0=\hbox{#1} \dp0=1.5pt \mathsurround=0pt
   \underline{\box0}$}   \def\ub{\underbar}    \def\ul{\underline}
\def\f{\left} \def\g{\right} \def\e{{\rm e}} \def\o{\over} \def\d{{\rm d}}
\def\vf{\varphi} \def\pl{\partial} \def\cov{{\rm cov}} \def\ch{{\rm ch}}
\def\la{\langle} \def\ra{\rangle} \def\EE{e$^+$e$^-$} \def\pt{p_{\rm t}}
\def\bitz{\begin{itemize}} \def\eitz{\end{itemize}}
\def\btbl{\begin{tabular}} \def\etbl{\end{tabular}}
\def\btbb{\begin{tabbing}} \def\etbb{\end{tabbing}}
\def\beqar{\begin{eqnarray}} \def\eeqar{\end{eqnarray}}
\def\\{\hfill\break} \def\dit{\item{-}} \def\i{\item}
\def\bbb{\begin{thebibliography}{9} \itemsep=-1mm}
\def\ebb{\end{thebibliography}} \def\bb{\bibitem}
\def\bpic{\begin{picture}(260,240)} \def\epic{\end{picture}}
\def\akgt{\noindent{Acknowledgements}}
\def\fgn{\noindent{\bf\large\bf figure captions}}
\def\lan{\langle}
\def\ran{\rangle}
\def\p{\pi}
\def\ifmath#1{\relax\ifmmode #1\else $#1$\fi}%
\def\rc{\ifmath{{\mathrm{c}}}}
\def\cut{\ifmath{{\mathrm{cut}}}}
\def\rF{\ifmath{{\mathrm{F}}}}
\def\rK{\ifmath{{\mathrm{K}}}}
\def\rp{\ifmath{{\mathrm{p}}}}
\def\rt{\ifmath{{\mathrm{t}}}}
\def\LAB{\ifmath{{\mathrm{LAB}}}}
\def\cut{\ifmath{{\mathrm{cut}}}}
\def\beq{\begin{equation}}
\def\eeq{\end{equation}}

\newcommand{\cinst}[2]{$^{\mathrm{#1}}$~#2\par}
\newcommand{\crefi}[1]{$^{\mathrm{#1}}$}
\newcommand{\crefii}[2]{$^{\mathrm{#1,#2}}$}
\newcommand{\crefiii}[3]{$^{\mathrm{#1,#2,#3}}$}
\newcommand{\HRule}{\rule{0.5\linewidth}{0.5mm}}

\bd

\title{Charge Fluctuations in $\pi^{+}\rp$ and $\rK^{+}\rp$ Collisions at 250 GeV/$c$}

\author{M.R.~Atayan}
\affiliation{ Institute of Physics, AM-375036 Yerevan, Armenia}

\author{Bai Yuting}
\affiliation{ Institute of Particle Physics, Hua-Zhong Normal
University, Wuhan 430070, China}

\author{E.A.~De Wolf}
\affiliation{ Department of Physics, Universitaire Instelling
Antwerpen, B-2610 Wilrijk, Belgium}

\author{A.M.F.~Endler}
\affiliation{ Centro Brasileiro de Pesquisas Fisicas, BR-22290 Rio
de Janeiro, Brazil}

\author{Fu Jinghua}
\affiliation{ Institute of Particle Physics, Hua-Zhong Normal
University, Wuhan 430070, China}

\author{H.~Gulkanyan}
\affiliation{ Institute of Physics, AM-375036 Yerevan, Armenia}

\author{R.~Hakobyan}
\affiliation{ Institute of Physics, AM-375036 Yerevan, Armenia}

\author{W.~Kittel}
\affiliation{ High Energy Physics Institute (HEFIN), University of
Nijmegen/NIKHEF, NL-6525~ED Nijmegen, The Netherlands}

\author{Liu Lianshou}
\affiliation{ Institute of Particle Physics, Hua-Zhong Normal
University, Wuhan 430070, China}

\author{Li Zhiming}
\affiliation{ Institute of Particle Physics, Hua-Zhong Normal
University, Wuhan 430070, China}

\author{Z.V.~Metreveli}
\affiliation{ Institute for High Energy Physics of Tbilisi State
University, GE-380086 Tbilisi, Georgia; now at Northwestern Univ.,
Evanston, U.S.A.}

\author{L.N.~Smirnova}
\affiliation{ Scobeltsyn Institute of Nuclear Physics, Lomonosow
Moscow State University, RU-119899 Moscow, Russia}

\author{L.A.~Tikhonova}
\affiliation{ Scobeltsyn Institute of Nuclear Physics, Lomonosow
Moscow State University, RU-119899 Moscow, Russia}

\author{A.G.~Tomaradze}
\affiliation{ Institute for High Energy Physics of Tbilisi State
University, GE-380086 Tbilisi, Georgia; now at Northwestern Univ.,
Evanston, U.S.A.}

\author{F.~Verbeure$^\dagger$}
\affiliation{ Department of Physics, Universitaire Instelling
Antwerpen, B-2610 Wilrijk, Belgium}

\author{Wu Yuanfang}
\affiliation{ Institute of Particle Physics, Hua-Zhong Normal
University, Wuhan 430070, China}

\author{S.A.~Zotkin}
\affiliation{ Scobeltsyn Institute of Nuclear Physics, Lomonosow
Moscow State University, RU-119899 Moscow, Russia}
\affiliation{Now at DESY, Hamburg, Germany}

\collaboration{EHS/NA22 Collaboration}

\begin{abstract}

We report on measurements of event-by-event charge fluctuations in
$\pi^{+}\rp$ and $\rK^{+}\rp$ collisions at 250 GeV/$c$. The
dependence of these fluctuations on the size of the rapidity
windows are presented for the first time in the full phase space
domain. The corrections for the influence of global charge
conservation and leading-particle stopping are tested by the data.
The discrepancy due to incomplete correction given by STAR and
PHENIX are estimated. The dependence of the fluctuations on the
position of the rapidity bin and on the multiplicity at different
rapidity windows are also presented.
\end{abstract}


\maketitle

The subject of event-by-event fluctuations has currently drawn a
lot of attention in both theoretical and experimental studies of
relativistic heavy ion collisions~\cite{reviews,oview}. It is
argued that information on the QCD phase transition---the
formation of Quark Gluon Plasma (QGP)---can be inferred from
measurements~\cite{qgp} among which event-by-event charge
fluctuations are considered as a promising
signature~\cite{charges,oview}. Due to the fractional electric
charges of quarks, the charges spread more evenly throughout the
QGP volume than in a hadronic gas and, therefore, the fluctuations
are expected to suffer an observable suppression in a
QGP~\cite{charges}.


Recently, it has been demonstrated that event-by-event charge
fluctuation can be directly related to a thermodynamic
signature---the anomalous proton-number fluctuation at the
critical point~\cite{hatta}, which is supposed to enhance the
charge fluctuations. The observed enhancement of charge
fluctuations at RHIC and SPS ~\cite{oview} seems to be a good
support for these arguments, though the effects of limited
detector acceptance and other corrections need to be further
investigated.

The charge fluctuations are also sensitive to other effects, as
the number of resonances at chemical freeze-out~\cite{koch, heinz}
and fluctuations occurring in the initial stage~\cite{marek}. The
corresponding analyses are interesting by their own beyond the QGP
hypothesis~\cite{jacek}.

There are mainly two kinds of measures for the event-by-event
charge fluctuations on the market at present, others being related
to these under reasonable assumptions~\cite{oview}. One is that of
{\it net charge } fluctuations, the other that of {\it charge
ratio} fluctuations. The direct measure of {\it net charge}
fluctuations is the variance of net charge $Q$, \beq \delta
Q^2=\langle Q^2\rangle -\langle Q\rangle^2, \qquad Q=n^+-n^-, \eeq

\noindent where $n^+$ and $n^-$ are the numbers of positive and
negative particles observed in a particular phase space window
under consideration. The average is over all events in the sample.
If charge is randomly assigned to each particle, $\delta
Q^2=\langle n_{\rm ch} \rangle$, where $n_{\rm ch}=n^++n^-$. So
the measure for net charge fluctuations is defined as \beq
D(Q)=4\frac{\delta Q^2}{\langle n_{\rm ch}\rangle}, \eeq \noindent
{\it i.e.}, equal to $4$ for independent emission.

In order to reduce the fluctuations of $n_+$ and $n_-$ due to the
variation of impact parameter, {\it charge ratio} $R=n^+/n^-$
fluctuations are recommended in \cite{charges} and the
corresponding measure is \beq D(R)=\langle n_{\rm ch}\rangle\cdot
\delta R^2, \eeq

\noindent where $\delta R^2=\langle R^2 \rangle -\langle
R\rangle^2$.

In the high multiplicity limit, the above two measures are
approximately equal, with the leading order correction being $\sim
1/\langle n_{\rm ch}\rangle$.

In accounting for the charge conservation in a large rapidity
window and a non-zero net charge due to non-negligible baryon
stopping, two correction factors~\cite{ccharge}, \beq
C_y=1-\frac{\langle n_{\rm ch}\rangle_{\Delta y}}{\langle n_{\rm
ch}\rangle_{\rm total}},
 \ \  C_{\mu}=\frac{\langle n_{\Delta y}^+\rangle^2}{\langle n_{\Delta y}^-\rangle^2} \ ,
\eeq \noindent are applied to the $D$-measures of Eq's.~(2) and
(3): \beq \tilde D=\frac{D}{C_yC_{\mu}}. \eeq

\begin{figure*}
\centering
\includegraphics[width=6.in]{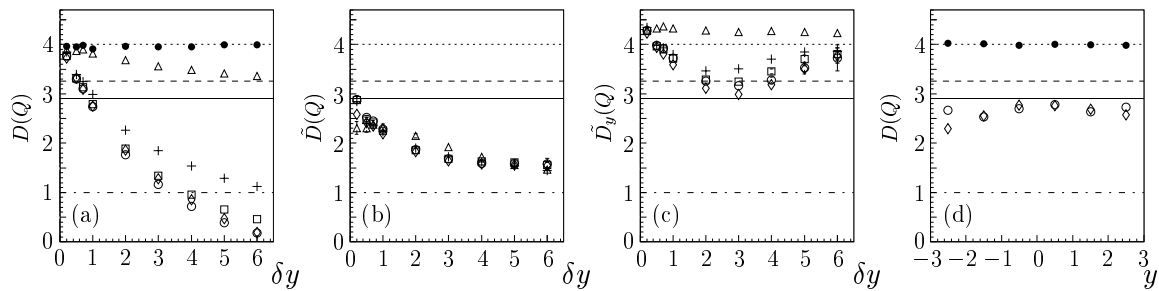}
\caption{\label{Fig. 1}(a)~Dependence of $D(Q)$ on the size of the
central rapidity window $|y|<\delta y/2$ with $0.001{\rm GeV}/c
<p_{\rm t}<10{\rm GeV}/c$ (open circles), $p_{\rm t}>0.1{\rm
GeV}/c$ (open squares), $p_{\rm t}>0.2{\rm GeV}/c$ (crosses), and
$p_{\rm t}>0.2{\rm GeV}/c,\  \Delta \phi=\pi/2$ (open triangles),
(b)~Corrected version of (a) by two factors. (c)~Corrected version
of (a) by only the $C_y$ factor of Eq.~(4). (d)~Dependence of
$D(Q)$ on the position of a unit-width rapidity window. Dotted,
dashed, solid and dot-dashed lines correspond to independent
emission, quark coalescence, resonance gas and QGP, respectively.
The solid circles correspond to random charge assigned to each
particle and open diamonds are the results from PYTHIA. }
\end{figure*}

The theoretical prediction~\cite{charges, oview} for the
$D$-measure in a QGP is $D=1$. It is 2.9 for a resonance
gas~\cite{charges, oview} and 3.26 in a quark coalescence
model~\cite{mquark, oview}.

Before comparing the data from different experiments with the
above expectations, one must know how the measurements depend on
the size of the rapidity window. This dependence has been
estimated in various models~\cite{charges, zaranek}, but the
results depend strongly on the assumptions for the rapidity
correlator and the width of acceptance. Therefore, a
model-independent study of the dependence of the fluctuations on
the size of the rapidity window in over the full rapidity domain
is called for.

In addition, one should test whether the correction factors given
by Eq.~(4) are valid. If correct, a rapidity window size scaling
should be observed in large rapidity windows due to the global
charge conservation. Moreover, the rapidity size for the onset of
the scaling will offer us a valuable scale for the relaxation time
of long-range correlation caused by charge
conservation~\cite{hatta, shuryak}.

Due to the limited acceptance in current heavy ion
experiments~\cite{phenix, star, na49}, this kind of study can only
be performed in hadron-hadron experiments, such as NA22, which is
equipped with a rapid cycling bubble chamber as an active vertex
detector and has excellent momentum resolution over its full
$4\pi$ acceptance.

In this letter, the dependence of the event-by-event net charge
and charge ratio fluctuations on the size of the rapidity window
is presented for $\pi^{+}\rp$ and $\rK^{+}\rp$ collisions at 250
GeV/$c$. Since no statistically significant differences are seen
between the results for $\pi^{+}$ and $\rK^{+}$ induced reactions,
the two data samples are combined for the purpose of this
analysis. A total of 44~524 non-single-diffractive events is
obtained after all necessary selections as described in detail
in~\cite{na22data}. Secondary interactions are suppressed by a
visual scan and the requirement of charge balance, $\gamma$
conversion near the vertex by electron identification.

The $D(Q)$-measure in central rapidity windows $|y|\hskip-1mm <
\hskip-1mm y_{\rm cut}$=$\delta y/2$ with $\delta y=0.2$, 0.5,
0.7, 1.0 to 6 is presented in Fig.~1(a), where the open circles
are the NA22 data and the open diamonds correspond to the results
from PYTHIA 5.720~\cite{pythia} (this convention will be kept in
all the following figures). In order to compare the results to
those from STAR and PHENIX, data for the same low $p_{\rm t}$ and
azimuthal cuts as STAR~\cite{star} and PHENIX~\cite{phenix} are
also presented.

The solid circles correspond to random charge assigned to each
particle, which indeed gives the value of $4$ as expected, no
matter how small the multiplicity is in very narrow rapidity
intervals. This shows that the accuracy of event-by-event analysis
hardly depends on event multiplicity and thus can be useful even
for low multiplicity cases~\cite{bialas}. So, the dependence of
the data on centrality is not caused by an insufficient number of
particles~\cite{stanislaw}.

\begin{table}
\caption{\label{Table 1.} $D(Q)$ in different phase-space
domains.}
\begin{tabular}{ccc}
\hline\small
   phase-space domain & RHIC & NA22 \\ \hline
  $|y|<0.35$, $p_{\rm t}>0.2$GeV/$c$,  &$ 3.86\pm 0.028$&  $3.896\pm 0.025$\\
  $\Delta \phi=\pi/2.$ & (PHENIX) & \\ \hline
 $|y|<0.5$, $p_{\rm t}>0.1$GeV/$c$ &$2.8\pm 0.05$(cent.) & $2.786\pm 0.015$ \\
 &$3.1\pm 0.05$(peri.)& \\
 &(STAR)&\\ \hline
\end{tabular}
\end{table}

From Fig.~1(a), it can be seen that with the widening of the
rapidity window, the NA22 data keep decreasing from the value
close to 4 (as expected for independent emission) to 1 (as
expected for a QGP) and even below. The loss of small-$p_{\rm t}$
particles and the cut in azimuthal angle both enhance the
fluctuations. The $D(Q)$ obtained under the same cuts as
PHENIX~\cite{phenix} and STAR~\cite{star} are listed in Tab.~I.
Their values are consistent with ours.

Taking into account charge conservation in large rapidity windows
and leading-particle stopping, the corrected measure $\tilde D(Q)$
is presented in Fig.~1(b). The results decrease from about $2.9$
(as expected for resonance gas) to above 1. The scaling appears
when the size of the central rapidity window is larger than
$|y|<2$, showing that the influence of charge conservation and
leading-particle stopping have been well eliminated by the factor
defined in Eq.~(4) and the correlation length of charge
conservation is about $4$ rapidity units. The corrections reduce
the measure in small rapidity windows and enhance it in large
ones. Since the influence of global charge conservation always
enhances the fluctuations, {\it i.e.}, $C_y<1$ in Eq.~(4), the
suppression in small rapidity windows shows that the
leading-particle stopping is non-negligible. If only the effect of
global charge conservation is taken into account, as in
STAR~\cite{star}, this will always enhance the fluctuations and
the scaling behavior disappears. The results for such a correction
are presented in Fig.~1(c). So, the data from both STAR and PHENIX
exaggerate the fluctuations, the former considering only one
correction and the latter without corrections at all.

In Fig.~1(d), $D(Q)$ is presented for different positions of a
unit-width rapidity window. It is almost a constant near that of a
resonance gas, showing that the charge fluctuations are
insensitive to the position of the rapidity window and that the
local charge is non-equilibrium, as pointed out in~\cite{wuliu}.

We now turn to a similar study of the charge ratio fluctuations.
Due to the positive charge of the initial-state particles, the
average number of positively charged particles is higher than that
of negatively charged ones. Therefore, we present the $D$-measures
in terms of the charge ratios $R^+=n^+/n^-$ and $R^-=n^-/n^+$ in
Fig's.~2(a) and (b) separately, where events with $n^-=0$ and
$n^+=0$ have been excluded from the analysis of $R^+$ and $R^-$,
respectively. It can be seen from the figures that $D(R^+)$ have
much larger values than $D(R^-)$. Both of them have behavior very
different from that of net charge fluctuations.

\begin{figure}
\includegraphics[width=2.4in]{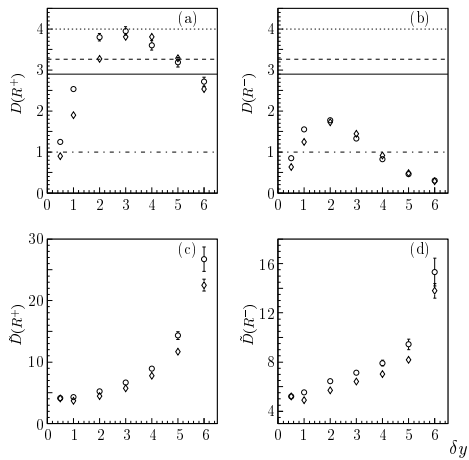}
\caption{\label{Fig. 2} The dependence of $D(R^+)$ and $D(R^-)$
(upper row) and their corrected versions (lower row) on the size
of the central rapidity window $|y|<\delta y/2$. Open circles are
the NA22 data and open diamonds are the results from PYTHIA.}
\end{figure}

The charge ratio measures corrected according to Eq.~(4) is given
in Fig's.~2(c) and (d). All points are above independent emission
and increase rapidly with the widening of the central rapidity
window, in analogy with the model calculation~\cite{jpciae} for
A-A collisions. These results show that the corrections proposed
for {\it net charge} in the observed window as given by Eq.~(4)
are invalid for {\it charge ratio} fluctuations.

It is further interesting to check how these measures do in
recording the {\it change} of charge fluctuations with
multiplicity in different rapidity windows. This is important, in
particular, because the even- and odd-multiplicity distributions
coincide in small rapidity windows, {\it e.g.} $|y|<2$, while
separation of them appears in large windows, {\it e.g.},
$|y|<3.$~\cite{na22n}.

\begin{figure}
\includegraphics[width=3.4in]{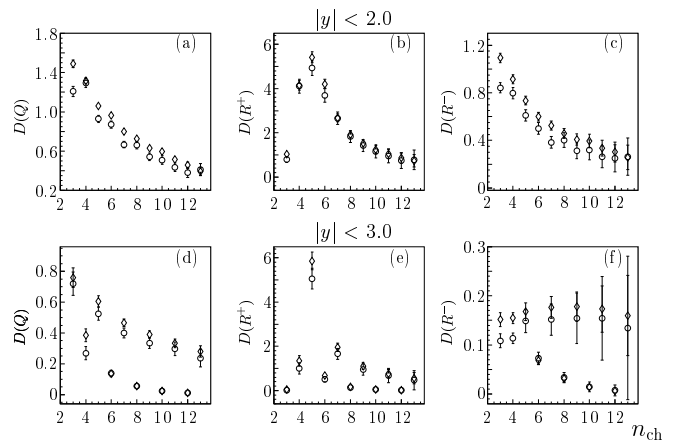}
\caption{\label{Fig. 3} The dependence of $D(Q)$, $D(R^+)$ and
$D(R^-)$ on multiplicity in $|y| < 2$ (1st row) and $|y|<3$ (2nd
row). Open circles are the NA22 data and open diamonds are the
results from PYTHIA.}
\end{figure}

The dependence of $D(Q)$, $D(R^+)$ and $D(R^-)$ on multiplicity in
two rapidity windows is presented in Fig's.~3. The following can
be observed: (1)~First of all, all plots show clear multiplicity
dependence, while the results from PHENIX~\cite{phenix} in a small
central rapidity window show that only $D(R)$ depends on
multiplicity, while $D(Q)$ is independent of it. (2)~For $|y|<2$,
the fluctuations of even and odd multiplicities in terms of net
charge and charge ratios coincide within the error bars,
consistent with the coincidence of even- and odd-multiplicity
distributions in small rapidity windows. (3)~For $|y|<3$, the
$D(Q)$ separate for even- and odd-multiplicities,  consistent with
the separation of even- and odd-multiplicity distributions in
large rapidity windows. The $D(Q)$ have almost equal separation
distance for all multiplicities. While the $D(R^+)$ and $D(R^-)$
are separated differently for different multiplicities, with very
big errors for odd multiplicities, as they could be the
combinations of very different $n^+$ and $n^-$. These observations
show that $D(Q)$ is better in recording the change of charge
fluctuations with multiplicity in different size of central
rapidity windows.

The above results can be summarized as follows: (1)~$D(Q)$,
$D(R^+)$ and $D(R^-)$ depend strongly on the size of the central
rapidity window. (2)~$\tilde D(Q)$ eliminates the influence of
global charge conservation and leading-particle stopping. Its
scaling behavior is observed when the central rapidity window is
wider than $4$ rapidity units. The same corrections are invalid
for charge ratio fluctuations. (3)~$D(Q)$ is insensitive to the
position of the rapidity bin. (4)~$D(Q)$, $D(R^+)$ and $D(R^-)$
all have clear multiplicity dependence. $D(Q)$ has a better record
in distinguishing the charge fluctuations of even and odd
multiplicities than $D(R^+)$ and $D(R^-)$. (5)~PYTHIA can
reproduce almost all the data for charge fluctuations, while it
fails to describe the transverse momentum fluctuations in
different central rapidity windows~\cite{na22phipt}.

In summary,  the dependence of charge fluctuations on the size of
the rapidity window is presented for the first time in the full
rapidity domain. The correction factors for net charge
fluctuations given by Eq.~(4) eliminate the influence of global
charge conservation and leading-particle stopping. The latter is
non-negligible in small rapidity windows. Due to the incomplete
consideration on these two corrections, both STAR and PHENIX
exaggerate the fluctuations. The scale of long-range correlations
caused by charge conservation is about $4$ rapidity units at
$\sqrt s =22$ GeV/$c$. The measure in terms of net charge
fluctuations is better than that of charge ratio ones.

This work is part of the research program of the "Stichting voor
Fundamenteel Onderzoek der Materie (FOM)", which is financially
supported by the "Nederlandse Organisatie voor Wetenschappelijk
Onderzoek (NWO)". We further thank NWO for support of this project
within the program for subsistence to the former Soviet Union
(07-13-038). The Yerevan group activity is financially supported,
in the framework of the theme No. 0248, by the Government of the
Republic of Armenia. This work is also supported in part by the
National Natural Science Foundation of China and the Ministry of
Education of China and by the Royal Dutch Academy of Sciences
under the Project numbers 01CDP017, 02CDP011 and 02CDP032 and by
the U.S. Department of Energy.

\ed


%